\documentclass[aps,pra,reprint]{revtex4-1}
\usepackage{amsmath,amssymb}
\usepackage{bm}
\usepackage{graphicx}
\graphicspath{{figures/}}
\usepackage{siunitx}
\usepackage{color}
\usepackage{hyphenat}
\usepackage[bookmarks=false]{hyperref}
\usepackage{braket}
\usepackage{multirow}
\usepackage{diagbox}
\usepackage{makecell}
\usepackage[capitalise]{cleveref}
\crefname{section}{Sec.}{Secs.}
\Crefname{section}{Section}{Sections}

\def\ket#1{\left| #1 \right\rangle}
\def\bra#1{\left\langle #1 \right|}

\newcommand{\beq}{\begin{equation}}
\newcommand{\eeq}{\end{equation}}

\definecolor{pink}{RGB}{255,0,255}

\begin{document}

\title{Implementation vulnerabilities in general quantum cryptography}

\author{Anqi~Huang$^{1,2,3}$}
\email{angelhuang.hn@gmail.com}
\author{Stefanie~Barz$^{4,5}$}
\author{Erika~Andersson$^{6}$}
\author{Vadim~Makarov$^{7,8}$}

\affiliation{
$^1$Institute for Quantum Information \& State Key Laboratory of High Performance Computing, College of Computer, National University of Defense Technology, Changsha 410073, People's Republic of China\\
$^2$Institute for Quantum Computing, University of Waterloo, Waterloo, ON, N2L~3G1 Canada\\
\mbox{$^3$Department of Electrical and Computer Engineering, University of Waterloo, Waterloo, ON, N2L~3G1 Canada}\\
\mbox{$^4$Institute for Functional Matter and Quantum Technologies, University of Stuttgart, 70569 Stuttgart, Germany}\\
\mbox{$^5$Center for Integrated Quantum Science and Technology, University of Stuttgart, IQST, 70569 Stuttgart, Germany}\\
$^6$Institute of Photonics and Quantum Sciences, Heriot-Watt University, Edinburgh, Scotland, EH14~4AS UK\\
$^7$Russian Quantum Center and MISIS University, Moscow, Russia\\
$^8$Department of Physics and Astronomy, University of Waterloo, Waterloo, ON, N2L~3G1 Canada
}

\date{\today}

\begin{abstract}
Quantum cryptography is information-theoretically secure owing to its solid basis in quantum mechanics. However, generally, initial implementations with practical imperfections might open loopholes, allowing an eavesdropper to compromise the security of a quantum cryptographic system. This has been shown to happen for quantum key distribution~(QKD). Here we apply experience from implementation security of QKD to several other quantum cryptographic primitives. We survey quantum digital signatures, quantum secret sharing, source-independent quantum random number generation, quantum secure direct communication, and blind quantum computing. We propose how the eavesdropper could in principle exploit the loopholes to violate assumptions in these protocols, breaking their security properties. Applicable countermeasures are also discussed. It is important to consider potential implementation security issues early in protocol design, to shorten the path to future applications.

\end{abstract}

\maketitle

\section{General introduction}

Common aims in cryptography are to guarantee confidentiality, integrity, and authentication of information. Some of the conventional cryptography based on computational complexity might be broken by a powerful quantum computer~\cite{ETSIwhitepaper}. However, quantum cryptography, where security rests on the laws of quantum mechanics, is one way to achieve information-theoretic security. Among the quantum cryptographic protocols, quantum key distribution (QKD) has become theoretically mature and technically practical. Inspired by the idea of QKD and taking advantage of QKD implementations, other quantum cryptographic primitives have gradually been developed, such as quantum coin tossing, quantum secret sharing, and quantum digital signatures~\cite{bennett1984,cleve1999,gottesman2001}. For each primitive, different protocols have been proposed, and even realized by current technology~\cite{bogdanski2008,collins2016,yin2017b}.

However, there is a non-negligible gap between theory and practice in QKD: imperfections in devices create various loopholes that compromise the protocol's security~\cite{zhao2008,lydersen2010a,weier2011,sun2011,jain2014,sajeed2015a,makarov2016}. Practical security issues might also occur in the realization of other quantum cryptographic protocols. In theory, the protocols are unconditionally secure, but the security might not be guaranteed in practice due to imperfections of devices. Investigating device imperfections and system loopholes in QKD has taken more than a decade, and is still in progress. The experience gained from QKD will be helpful in finding possible loopholes in other implementations of quantum cryptographic protocols, because they use similar optical components. This enhances the practical security of quantum cryptography.

\begin{table*}
  \caption{Summary of potential attacks in implementations of quantum cryptographic protocols. The table lists broken security properties for five primitives: two different protocols for quantum digital signatures (QDS), two different protocols for quantum secret sharing (QSS), source-independent quantum random number generation (SI~QRNG), quantum secure direct communication (QSDC), and blind quantum computing (BQC). ``--" means the attack is not applicable. See text for details.\\}
  \label{tbl:summary}
    \begin{tabular}{l|cccc}
      \diagbox{\bf{Protocol}~~~~}{\bf{Attack}~} & \makecell{Source\\ side channel} & \makecell{Wavelength-dependent\\ attack} & \makecell{Detector control\\ attack} & \makecell{Trojan-horse\\ attack} \\
       \hline \\[-1.9ex]
       QDS& &  &  & \\
       Identical-state-sharing~\cite{yin2016c} & Unforgeability & Unforgeability& Unforgeability& --\\
       Different-state-sharing~\cite{collins2016} & -- & --& Unforgeability& Unforgeability\\
       \\[-1.4ex] 
       QSS& &  &  & \\
       Entanglement-based~\cite{chen2005} & -- &--&Confidentiality&--\\
       Single-qubit~\cite{bogdanski2008} & -- &--&-- &Confidentiality\\
       \\[-1.4ex] 
       SI QRNG~\cite{cao2016} & -- & Randomness &  Randomness & -- \\
       \\[-1.4ex] 
       QSDC~\cite{hu2016}& --&--&Confidentiality&Confidentiality\\
       \\[-1.4ex] 
       BQC~\cite{barz2012}& Confidentiality&--&--&Confidentiality
  \end{tabular}
\end{table*}

The vulnerability of quantum coin-tossing and non-loophole-free Bell testing has previously been demonstrated \cite{sajeed2015,gerhardt2011a}, using imperfections in their specific experimental implementations to remove the protocol's quantum advantages. In this Article, we survey five quantum cryptographic primitives as examples to investigate practical security threats in their implementation. The primitives are quantum digital signatures (QDS), quantum secret sharing (QSS), source-independent quantum random number generation (SI QRNG), quantum secure direct communication (QSDC), and blind quantum computing (BQC). Based on attacks known to exist for QKD, we propose potential attacks on these primitives. The attacks may compromise the security of practical quantum cryptographic systems, without making the legitimate participants abort the cryptographic protocols. We summarize potential imperfections and broken security properties for all five primitives in~\cref{tbl:summary}. Details for each primitive are explained in~\cref{sec:QDS,sec:QSS,sec:QRNG,sec:QSDC,sec:BQC}. Each of these sections contains two parts: in subsection~A we recap the protocol, and in subsection~B we propose the attacks on its implementation. Countermeasures are discussed in~\cref{Countermeasure}. We conclude in~\cref{conclusion}. Please note that this study is merely a starting point presenting a broad overview. Detailed analysis of each implementation imperfection should be done in the future, as technological implementations of the protocols mature.

In this Article, we focus on the implementation security of the demonstrations. We also remark that while most of these quantum cryptographic schemes have advantages over ``classical'' schemes, for some of these protocols, their practical usefulness is less clear, and strict security proofs may still be under development. For example, it is not always clear what practical advantage all protocols for QSS offer, over protocols based on secret shared keys followed by a ``classical'' protocol for secret sharing. Similarly, for QSDC, one would need to motivate the usefulness of direct communication, as opposed to establishing secret shared keys using standard QKD, followed by encryption using these keys. Discussing these aspects is however outside the scope of our study.

\section{Quantum digital signatures}
\label{sec:QDS}

Digital signatures are an important primitive in cryptography. Specifically, three security properties are required for signatures: unforgeability, nonrepudiation, and transferability~\cite{swanson2011}. Unforgeability guarantees a unique message signer, so no one else is able to forge a valid signature. Nonrepudiation requires that once a message is signed, the signer cannot deny the signature. Transferability means that a recipient who accepts a message can be sure that if the message is forwarded, another recipient will also accept the message, except with a probability that can be made arbitrarily low. QDS based on laws of quantum physics is able to satisfy these requirements, and achieve information-theoretic security~\cite{gottesman2001}. Unconditionally secure signatures are also possible based on shared secret keys~\cite{chaum1990,hanaoka2000, swanson2011,amiri2016a}, and the scaling of secret key length with respect to message length can be more favorable than for quantum signatures. The secret shared key could be generated by quantum key distribution, but otherwise these schemes remain entirely ``classical''. On the other hand, the error rate threshold can be less strict for quantum digital signatures than for quantum key distribution to distill shared secret keys~\cite{amiri2016}.

References~\onlinecite{yin2016b, amiri2016} propose QDS protocols via insecure quantum channels, which later have been implemented~\cite{yin2016c, collins2016}. A significant difference between these two protocols is the stage of quantum state distribution. In Ref.~\onlinecite{yin2016b}, Alice sends the same quantum states to Bob and Charlie, while in Ref.~\onlinecite{amiri2016}, Bob and Charlie individually send different quantum states to Alice. 

Both protocols are briefly introduced in the next subsection. A reader familiar with protocol implementation can, of course, skip to subsection~B, where we discuss vulnerabilities.

\subsection{Protocol and implementation}

\subsubsection{Identical-state-sharing}
\label{ISS_QDS}

\begin{figure}
    \includegraphics{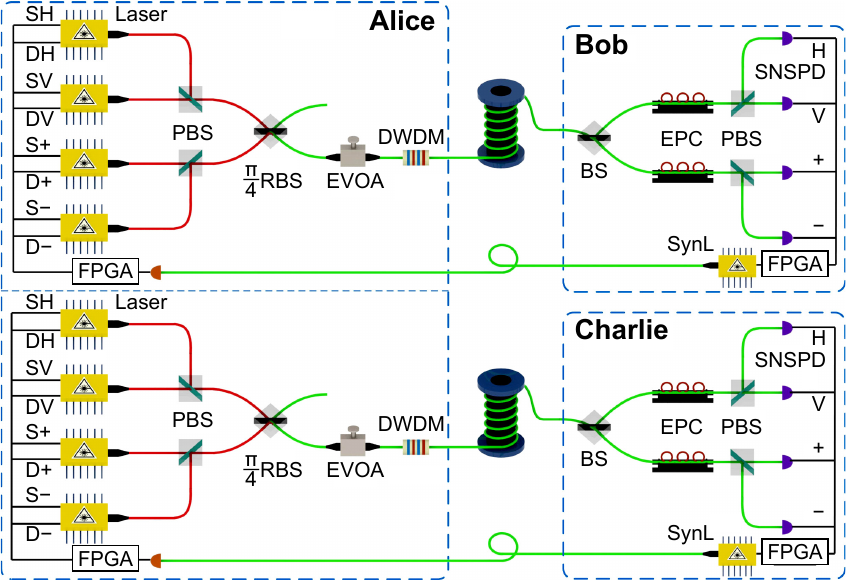}
    \centering
    \caption{Experimental setup for QDS implemented by H.-L. Yin and his coworkers (reprinted from Ref.~\onlinecite{yin2016c}). Alice first prepares two pairs of pairs of horizontally (H) and vertically (V) polarized photons using two pairs of lasers followed by polarization beam splitters~(PBS). One pair of H and V polarized photons are then rotated $\pi/4$ by a $\pi/4$ rotation beam splitter~($\frac{\pi}{4}$ RBS), becoming states $\ket{+}$ and $\ket{-}$. The variation in amplitude for the decoy states is implemented by an electrical variable optical attenuator~(EVOA). Bob and Charlie each randomly choose one of two bases to detect the incoming states. Here, DWDM denotes a dense wavelength division multiplexer, BS denotes a beam splitter, EPC denotes an electric polarization controller, SNSPD denotes a superconducting nanowire single-photon detector, SynL denotes a synchronization laser, FPGA denotes a field programmable gate array.}
    \label{fig:Yin_imp}
\end{figure}

Reference~\onlinecite{yin2016b} proposes a QDS protocol with a quantum-state sender, Alice, and two quantum-state receivers, Bob and Charlie. This protocol has been implemented over a distance of 102~km~\cite{yin2016c} as shown in~\cref{fig:Yin_imp}. The protocol consists of two stages, a quantum stage, and a signing stage. In the quantum stage, for each future 1-bit message $m = 0$ or $m = 1$, Alice employs weak coherent states to randomly prepare two identical sequences of qubit states, and every individual state is in one of the Bennett-Brassard 1984 (BB84) polarization states $\ket{H}$, $\ket{V}$, $\ket{+}$ and $\ket{-}$~\cite{bennett1984}. In addition, the decoy-state protocol~\cite{ma2005} is used to randomly modulate the mean photon numbers of the weak coherent states, protecting the system from photon-number-splitting~(PNS) attacks~\cite{brassard2000}. Then one copy of the sequence is sent to Bob, and one copy to Charlie. A beam splitter is used to randomly and independently select the $X$ or $Z$ basis to measure the received states.

In a sifting phase, Bob and Charlie announce in which slots they obtain detections. For each detection slot, Alice then announces two nonorthogonal states from different bases, for example, $\ket{H}$ and $\ket{+}$. One of them is the real state she sent. If Bob (Charlie) obtains a measurement result corresponding to a state that is orthogonal to one of the states Alice announced, such as $\ket{V}$, then Bob (Charlie) conclusively knows that it is the other announced state, $\ket{+}$.

In the next stage, the signing stage, only classical processing takes place. It starts by announcing some of the states shared between Alice and Bob (Charlie) during the quantum stage to calculate an authentication threshold $T_{a}$ ($T_{v}$) for Bob (Charlie). The unannounced states form strings denoted as $S_{Am}$ for Alice, $S_{Bm}$ for Bob and $S_{Cm}$ for Charlie, and will be used for the digital signature. To send a signed 1-bit message $m$, Alice sends the message and the corresponding data string, ($m$, $S_{Am}$), to one of the recipients, say, Bob. Bob will accept this message if the mismatch rate of sifted bits between $S_{Am}$ and $S_{Bm}$ is less than $T_{a}$. If Bob wishes to forward the message to Charlie, he forwards ($m$, $S_{Am}$) to Charlie. Charlie will accept this message as well if the mismatch rate of the sifted bits between $S_{Am}$ and $S_{Cm}$ is less than $T_{v}$.

\subsubsection{Different-state-sharing}
\label{DSS_QDS}

\begin{figure}
    \includegraphics{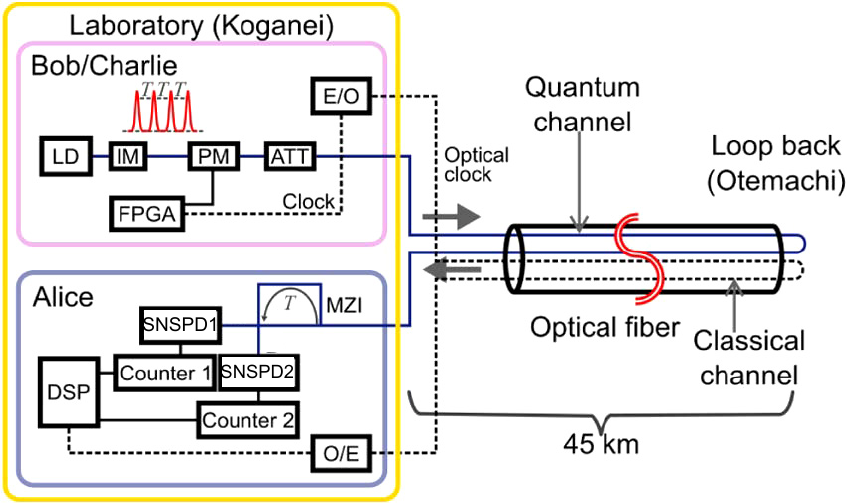}
    \centering
    \caption{Implementation of QDS by R. J. Collins and his coworkers, employing a DPS QKD system (reprinted with permission from Ref.~\onlinecite{collins2016}), (OSA). Bob and Charlie are the quantum-state transmitters, and Alice measures the received states. Here, LD is a laser diode, IM is an intensity modulator, PM is a phase modulator, ATT is an attenuator, FPGA is a field programmable gate array, E/O and O/E are electrical-to-optical and optical-to-electrical converters, SNSPD is a superconducting nanowire single-photon detector, DSP is a digital signal processor, MZI is a Mach-Zehnder interferometer.}
    \label{fig:Collins_imp}
\end{figure}

Reference~\onlinecite{amiri2016} proposes another quantum digital signature protocol that sends different quantum states from Bob and Charlie to Alice. This protocol has subsequently been implemented based on an installed differential-phase-shift (DPS) QKD system, as shown in~\cref{fig:Collins_imp}~\cite{collins2016}. This protocol is also divided into two stages, a distribution stage and a messaging stage. In the distribution stage, Bob and Charlie randomly and independently select two different \textit{n}-bit strings. Then, they encode the bits into quantum states according to the DPS QKD protocol~\cite{inoue2002}. For each future message $m = 0$ or $m = 1$, Bob (Charlie) applies a key-generating protocol (KGP) to share the bit string with Alice. The KGP can be treated as a partial QKD procedure without error correction and privacy amplification. Alice and Bob (Charlie) estimate the quantum bit error rate~(QBER) by announcing a small part of the shared bits. The remaining $L$-bit key is denoted by $K_{m}^{B}$ ($K_{m}^{C}$) at Bob's (Charlie's) side. At Alice's side, she obtains a signature $\textrm{Sig}_{m} = (A_{m}^{B}, A_{m}^{C})$ for a future message $m$. Then, to guarantee transferability, Bob and Charlie randomly forward half of their keys, $K_{m}^{B}$ and $K_{m}^{C}$, to each other. This classical bit exchange is encrypted by Bob and Charlie using a separate BB84 QKD system. This way, Alice receives no information on which bits have been forwarded and which bits have been kept. From her point of view, a bit she originally shared with Bob (Charlie) is now equally likely to be retained by Bob as by Charlie. Bob (Charlie) combine the non-exchanged part of $K_{m}^{B}$ ($K_{m}^{C}$) and the received part of $K_{m}^{C}$ ($K_{m}^{B}$) as a symmetric key, $S_{m}^{B}$ ($S_{m}^{C}$).

In the messaging stage, Alice signs a message $m$ by $\textrm{Sig}_{m}$, and then sends $(m, \textrm{Sig}_{m})$ to Bob. Bob checks the mismatch rate between $\textrm{Sig}_m$ and $S_m^B$. If the mismatch rate is lower than the threshold $s_a$, Bob accepts the message. If Bob wishes to forward the message to Charlie, he forwards $(m, \textrm{Sig}_{m})$ to Charlie. Charlie also checks the mismatch rate between $\textrm{Sig}_m$ and $S_m^C$, and accepts the message if the mismatch rate is lower than the threshold $s_v$. From Alice’s point of view, the situation is symmetric with respect to Bob and Charlie, so that if Bob accepts a signature, Charlie must accept it with high probability, provided acceptance thresholds are chosen correctly and differently for Bob and Charlie.

\subsection{Hacking}

Both protocols have been proven to be information-theoretically secure, based on different assumptions~\cite{yin2016b, amiri2016}. In this section, we analyse the security assumptions for both protocols and illustrate how these assumptions might be broken. Since QDS realizations are based on QKD schemes with similar optical components, similar vulnerabilities exist. That is, some known attacks on QKD systems are applicable also to the realisations of quantum digital signatures. In our analysis, we assume an external attacker Eve who is not a legitimate participant (Alice, Bob or Charlie) in the QDS protocol.

\subsubsection{Identical-state-sharing protocol}
\label{hack_ISS_QDS}

The unforgeability of this protocol is based on the assumption that given two copies of quantum states, Eve cannot distinguish between all four states Alice might send without error before Alice's declaration~\cite{chefles2001}. However, in practice, if Eve were able to discriminate the states via a side channel, messages could be forged. Several side channels exist in the implementation~\cite{yin2016c}, which could be exploited by Eve to hack it.

Source side channels are useful for Eve to learn the quantum state prepared by Alice. When quantum states are prepared by different laser diodes, side channels could exist both in time and frequency domains~\cite{nauerth2009,huang2018}. In the implementation presented in Ref.~\onlinecite{yin2016c}, each laser diode prepares a specific state, and different laser diodes are used in a random order. To avoid the spectral side channel, the implementation controls the difference of the central wavelengths for all of these laser diodes in a narrow range (0.02 nm). Additionally, a dense wavelength division multiplexer~(DWDM) with 100~GHz bandwidth is used as a filter before the states are sent out. However, a side channel might exist in another degree of freedom. For example, pulse emission time, pulse width and pulse shape may vary for different laser diodes. These mismatches give Eve a chance to distinguish different states~\cite{huang2018}. If Eve is able to perfectly distinguish the quantum states, she could forge a copy of Alice's signature and send it to Bob and Charlie. However, usually, Eve can only partially distinguish the states. She may choose to perform different types of quantum measurements to maximize her distinguishability. For example, if Eve makes a measurement that sometimes gives her higher confidence in the result, such as an unambiguous quantum measurement, then she could forward a state only when her measurement has succeeded. Thus, in this case, if losses are high enough, this strategy may not be noticed by the legitimate parties.

Measurements are usually more vulnerable than state preparation. One potential flaw hides in the beam splitter situated at the input of Bob's/Charlie's subsystem. The output ratio of the beam splitter might depend on the wavelength of the incoming light~\cite{li2011a}, which helps Eve during the intercept-resend attack. Eve first measures a state sent by Alice. According to the measurement result, Eve resends the measured state with a wavelength that makes the output ratio of the beam splitters become highly unbalanced, for example, 99:1 or 1:99. Then the resent state passes through Bob's/Charlie's beam splitter via one output with high probability, likely reaching the same measurement basis as Eve's. Thus, Eve, Bob and Charlie share almost the same detection results. At the sifting phase, Eve can wiretap the public announcement and follow the sifting rule described in the protocol, obtaining her signature string. After that, if the mismatch rate between Eve's and Bob's (Charlie's) strings is lower than $T_a$ ($T_v$), Eve would be able to pretend to be Alice and send a signature to Bob (Charlie).

To force Bob and Charlie to obtain the same detection results as Eve during the intercept-resend attack, another possible tool is a detector blinding attack~\cite{lydersen2010a, lydersen2011c, tanner2014}. By applying this attack, Eve might be able to control all Bob's measurement results~\cite{lydersen2010a, lydersen2011c, tanner2014}. In this attack, Eve sends a strong laser to blind Bob's and Charlie's detectors such that they are no longer sensitive to single photons, but act as classical optical detectors. Then, during intercept-resend, Eve resends the measured states by energy-tailored pulses. The resent pulses trigger Bob's detections in the same basis and state as Eve's. If the detector blinding attack is possible in this QDS implementation, Eve could obtain a copy of the bits shared by Alice and Bob/Charlie after sifting. Thus, Eve could pretend to be Alice to sign a message. The detector blinding attack can maintain the normal detection statistics~\cite{gerhardt2011}. Furthermore, in a receiver that uses a beam splitter to passively choose bases and is vulnerable to the detector blinding attack, Eve can force a click with $100\%$ probability~\cite{gerhardt2011}. If the digital signature scheme is built using detectors other than superconducting nanowire single-photon detectors used in the implementation shown in~\cref{fig:Yin_imp}, other types of detector-control attacks may also apply, such as efficiency mismatch~\cite{makarov2006}, after-gate~\cite{wiechers2011}, superlinearity~\cite{lydersen2011b}, and deadtime~\cite{weier2011}.

\subsubsection{Different-state-sharing protocol}
\label{hack_DSS_QDS}

In this protocol, unforgeability is based on the security of the KGP that guarantees $d(A_{i}^{B}, K_{i}^{B}) < d(E_\textrm{guess}, K_{i}^{B})$ (with high probability), where $d$ is the Hamming distance and $E_\textrm{guess}$ is Eve's attempt at guessing $K_{i}^{B}$~\cite{amiri2016}. However, this property could be broken as well, if Eve can learn the states sent by Bob (Charlie) or forces Alice to detect the same result as hers. Similar to the previous protocol, the implementation might also contain several loopholes.

Alice's SNSPDs might be vulnerable to the detector blinding attack~\cite{lydersen2010a, lydersen2011c, tanner2014}. Similarly to the previous implementation, the SNSPDs might be blinded by a strong laser. Eve then does intercept-resend and sends Alice faked states whose power and phase are tailored~\cite{lydersen2011}. Thus, Eve, Alice and Bob (Charlie) share the same bit string, which means $d(A_{i}^{B}, K_{i}^{B}) = d(E_\textrm{guess}, K_{i}^{B})$.

At the source in Bob (Charlie), all the states are modulated by a phase modulator, which might open another loophole. The modulation information from the PM could be eavesdropped by a Trojan-horse attack~\cite{vakhitov2001,gisin2006, jain2014, sajeed2015}. In this attack, Eve sends strong light to Bob (Charlie). The reflected light carries the modulation information, which could be measured from the phase difference between injected light and reflected light. It has been shown that around four reflected photons are sufficient to read out most of the information~\cite{jain2014}. If the Trojan-horse attack is successful in the QDS system, Eve could get all Alice's information: $d(E_\textrm{guess}, K_{i}^{B})$ could become equal to $d(A_{i}^{B}, K_{i}^{B})$.

\section{Quantum secret sharing}
\label{sec:QSS}

In secret sharing protocols, information is shared among many parties. The information can be reconstructed only if groups of parties collaborate. Information-theoretically secure secret sharing is possible not only using classical means (e.g., by pairwise shared keys), but also using quantum methods. Here, we focus on quantum secret sharing~\cite{cleve1999}. Two types of quantum secret sharing schemes, entanglement-based schemes~\cite{chen2005} and single-qubit schemes~\cite{bogdanski2008}, have been proposed for the sharing of classical messages. We survey both schemes.

\subsection{Protocol and implementation}

\subsubsection{Entanglement-based protocol}

\begin{figure}
    \includegraphics[width=0.85\columnwidth]{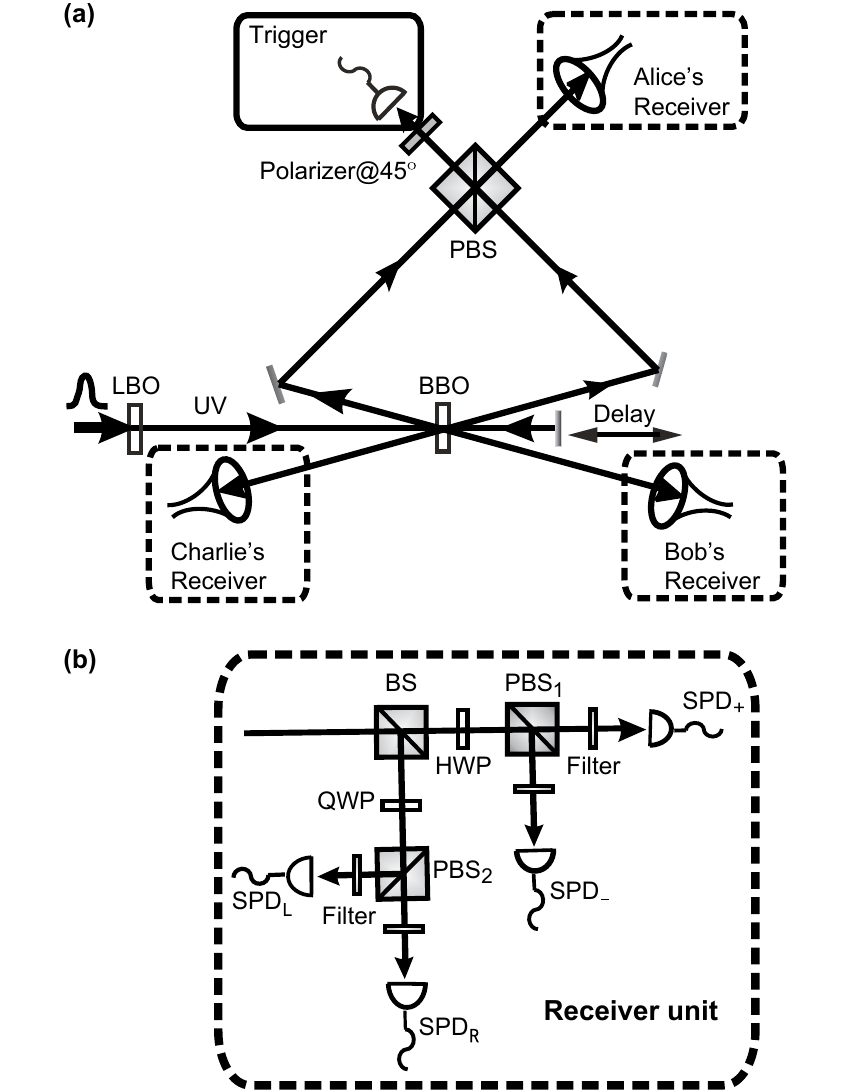}
    \caption{QSS based on entangled states (reprinted from Ref.~\onlinecite{chen2005}). (a) QSS system scheme. Ultraviolet (UV) pulses with a central wavelength of 394~nm are generated by a $\text{LiB}_3\text{O}_5$ (LBO) crystal. The pulses pass a beta-barium borate (BBO) crystal twice to generate two pairs of entangled photons. A photon triggers detection to synchronize GHZ state detections at Alice, Bob and Charlie. (b) Optical structure of each receiver unit. Here, BS is a beam splitter, HWP is a half-wave plate, QWP is a quarter-wave plate, PBS is a polarization beam splitter, SPD is a single-photon detector.}
    \label{fig:entangle_setup}
\end{figure}

In one scheme for entanglement-based QSS~\cite{chen2005}, Alice, Bob and Charlie first hold one photon each in a Greenberger-Horne-Zeilinger (GHZ) triplet, which is the state
\begin{equation}
\label{eq:state}
\ket{\psi}_{ABC}= \frac{1}{\sqrt{2}}(\ket{H}_{A}\ket{H}_{B}\ket{H}_{C} + \ket{V}_{A}\ket{V}_{B}\ket{V}_{C}).
\end{equation}
Then, a projective measurement is performed on each photon randomly either in the $X$ or $Y$ basis, where the basis states are given by
\begin{equation}
\ket{X_{\pm}}= \frac{1}{\sqrt{2}}(\ket{H}\pm\ket{V}),  ~\ket{Y_{\pm}}= \frac{1}{\sqrt{2}}(\ket{H}\pm i\ket{V}).
\end{equation}
The GHZ states can be written as
\begin{eqnarray}
\label{eq:measured state}
\ket{\psi}_{ABC}&=\frac{1}{2}[(\ket{X_+}_{A}\ket{X_+}_{B} + \ket{X_-}_{A}\ket{X_-}_{B})\ket{X_+}_{C}\nonumber \\
&+ (\ket{X_+}_{A}\ket{X_-}_{B} + \ket{X_-}_{A}\ket{X_+}_{B})\ket{X_-}_{C}].
\end{eqnarray}
Thus, if each party measures in the $X$ basis, the measurement results would show perfect correlations. Once any two measurement results are known, the third measurement result can be predicted with certainty. Similar correlation would be obtained for three other measurement combinations, $X_A Y_B Y_C$, $Y_A X_B Y_C$, and $Y_A Y_B X_C$. However, the remaining four basis combinations, $X_A X_B Y_C$, $X_A Y_B X_C$, $Y_A X_B X_C$, and $Y_A Y_B Y_C$, result in uncorrelated measurement results among the three parties. Thus, they could announce their basis choices to sift the basis combinations with perfect correlation. After that, Alice and Bob share their measurement results with each other to establish Charlie's key. Thus, a message encrypted by Charlie can be decrypted if Alice and Bob cooperate. The protocol implementation is shown in~\cref{fig:entangle_setup}.

\subsubsection{Single-qubit protocol}
\label{QSS_single}

Instead of using entangled states, reference~\onlinecite{Schmid2005} proposed an \textit{N}-party QSS protocol that uses a single qubit, which is easily realizable and scalable compared to the entanglement-based protocol. On the other hand, this protocol completely removes the possibility to share quantum information in terms of an entangled state. The information shared is necessarily classical. Reference~\onlinecite{bogdanski2008} demonstrated this protocol. An initial qubit $\ket{x}= (\ket{0}+\ket{1})/\sqrt{2}$ is prepared by party $R_1$, and sent from $R_2$ to $R_N$ sequentially. Each party $R_i$ (\textit{i} = 1, ...,  $N-1$) encodes information by applying a phase randomly chosen from two sets, \{0, $\pi$\} and \{$\pi/2$, $3\pi/2$\}, to the $\ket{1}$ component in the qubit $\ket{x}$. The party $R_N$ randomly applies phase 0 or $\pi/2$ to the $\ket{1}$ component before measuring the state $\ket{\pm x} = (\ket{0}\pm \ket{1})/ \sqrt{2}$. Thus, the detection probability of each detector is
\begin{equation}
\begin{split}
P_{D1}= \frac{1}{2}[1+\cos(\sum_{i = 1}^N\phi_i)],\\
P_{D2}= \frac{1}{2}[1-\cos(\sum_{i = 1}^N\phi_i)].
\end{split}
\end{equation}

\begin{figure}
    \includegraphics[width=\columnwidth]{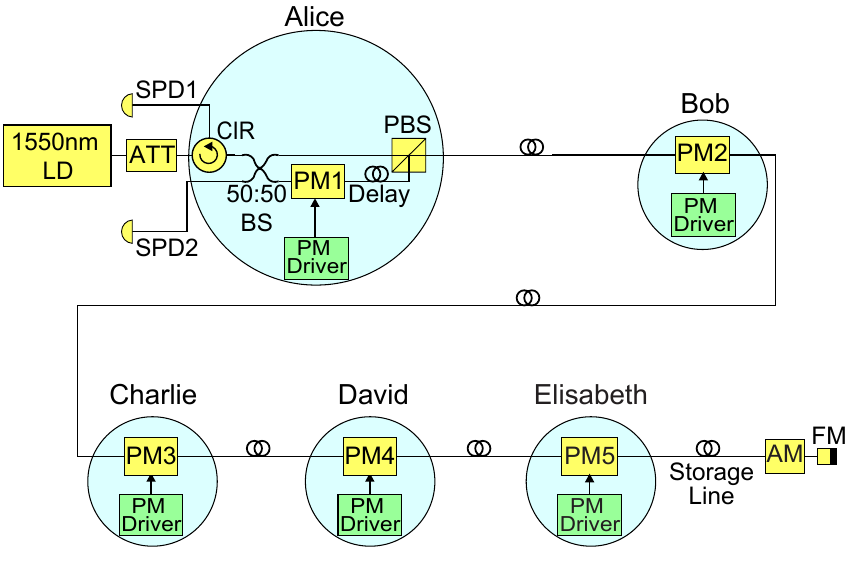}
    \caption{Single-qubit QSS (reprinted from Ref.~\onlinecite{bogdanski2008}). Alice randomly modulates the state, adding a phase of $0$ or $\pi/2$. The rest of parties randomly choose a phase from \{$0$, $\pi/2$, $\pi$, $3\pi/2$\}. Here, LD denotes a laser diode, ATT denotes an attenuator, SPD denotes a single-photon detector, CIR denotes a circulator, BS denotes a beam splitter, PBS denotes a polarization beam splitter, PM denotes a phase modulator, AM denotes an amplitude modulator, FM denotes a Faraday mirror.}
    \label{fig:single_setup}
\end{figure}

Half of the time, there is destructive or constructive interference, when $\cos(\phi_1 + \cdots + \phi_N) = \pm 1$. If all parties announce which set of phase values their choice belonged to, then every party knows which detection results are deterministic. Using the knowledge of which measurement results are deterministic, multiple parties can then share a secret as follows. If any $N-1$ parties collaborate and share their modulating phases, they would be certain about the phase applied by the $N$th party for one slot of the deterministic measurement. To maintain stability in the experiment, a bidirectional scheme is applied to implement a 5-party protocol~\cite{bogdanski2008} as shown in~\cref{fig:single_setup}. Alice prepares the initial pulse without phase encoding and acts as $R_5$ to measure the final reflected state. The rest of parties encode their information on the way back from the Faraday mirror~(FM), after the pulse is attenuated to the single-photon level by the amplitude modulator~(AM). This idea is similar to the plug~\&~play QKD system~\cite{muller1997}.

\subsection{Hacking}

We discuss one type of known attack that may work for the implementation of each aforementioned QSS protocol. An external Eve is assumed to be the attacker. If an external Eve can compromise the security, any inside attacker (a protocol participant) could also compromise security and obtain the secret without the cooperation of the other participants, because inside attackers have at least as much information as an outside attacker.

\subsubsection{Blinding attack on entanglement-based implementation}

In the entanglement-based QSS scheme mentioned above, three parties securely share a secret string using a GHZ state that has inherent correlations among the three photons. If Eve would like to perform an intercept-resend attack via a quantum channel, she would break the initial correlation between the three entangled photons, and thus introduce errors~\cite{hillery1999}. However, the detector blinding attack~(see~\cref{hack_ISS_QDS}) could help Eve steal the shared secret while introducing no error. Eve performs two independent detector blinding attacks on Alice's and Bob's detectors. The blinded detectors only click when Alice/Bob chooses the same measurement bases as Eve during an intercept-resend attack. Thus, Alice's and Bob's secret strings could be obtained by Eve to let her learn Charlie's key. Alternatively, instead of hacking Alice and Bob, Eve can directly blind Charlie's detectors to control the secret key he obtains.

\subsubsection{Trojan-horse attack on single-qubit implementation}
\label{two_way_trojan}

The security of single-qubit QSS follows the proven BB84 QKD protocol~\cite{Schmid2005}. Similarly to BB84 protocol, an intercept-resend attack on the QSS introduces $25 \%$ error in the final detection results. However, the implementation might have side channels that leak information about state preparation, allowing Eve to learn the shared secret without disturbing the normal QSS protocol.

In the implementation scheme of single-qubit QSS shown in~\cref{fig:single_setup}, similar to QKD systems, the phase modulation is implemented by a phase modulator which may be vulnerable. Thus, the Trojan-horse attack~(see \cref{hack_DSS_QDS}) appears to be a high risk, owing to the pass-through nature of every party except for Alice. Eve could send strong light to each party, excluding Alice, and then at the other side of each party receive the light modulated by the PM. By measuring the phase difference between Eve's original coherent light and the modulated light, she could read the phase modulation. In this way, Eve could know the secret shared among the four parties. In general, this hacking strategy works for $N$ parties. An attack on Alice may also be attempted, however, it is more difficult owing to the presence of SPDs in Alice~\cite{sajeed2017}.

\section{Source-independent quantum random number generation}
\label{sec:QRNG}

Quantum random number generation~(QRNG) based on the uncertainty principle in quantum mechanics can be used to provide pure random numbers, a crucial resource in cryptography~\cite{herrero2017}. Similarly to QKD, a QRNG system also consists of quantum state preparation and measurement, however, the states are measured locally without long-distance transmission. Source-independent~(SI) QRNG protocols~\cite{vallone2014, cao2016, marangon2017} assume that the state preparation setup is untrusted, while the measurement setup is trusted. We survey the protocol in Ref.~\onlinecite{cao2016} and its experimental demonstration.

\subsection{Protocol and implementation}

In the SI~QRNG protocol, an untrusted party Eve prepares single-qubit states $\ket{+}$ and sends them to Alice's measurement station~\cite{cao2016}. Alice first projects the quantum states into qubits $\ket{+}$ and vacuum states, but it is unclear how to implement this operation in practice. Assume that $n$ squashed qubits are obtained during the operation of the protocol. The resulting single qubits are then randomly measured either in the $X$ basis, \{$\ket{+}$, $\ket{-}$\}, or the $Z$ basis, \{$\ket{H}$, $\ket{V}$\}. If $n_x$ out of the $n$ squashed qubits are measured in the $X$ basis, they should be detected as $\ket{+}$ ideally. The detection rate for $\ket{-}$ is treated as the estimated error rate $e_{bx}$. The remaining $n_z=n - n_x$ qubits are measured in the $Z$ basis to generate $n_z$ random bits. Alice then extracts the final secure random bits from $n_z$, which is equivalent to privacy amplification in QKD.

\begin{figure}
    \includegraphics{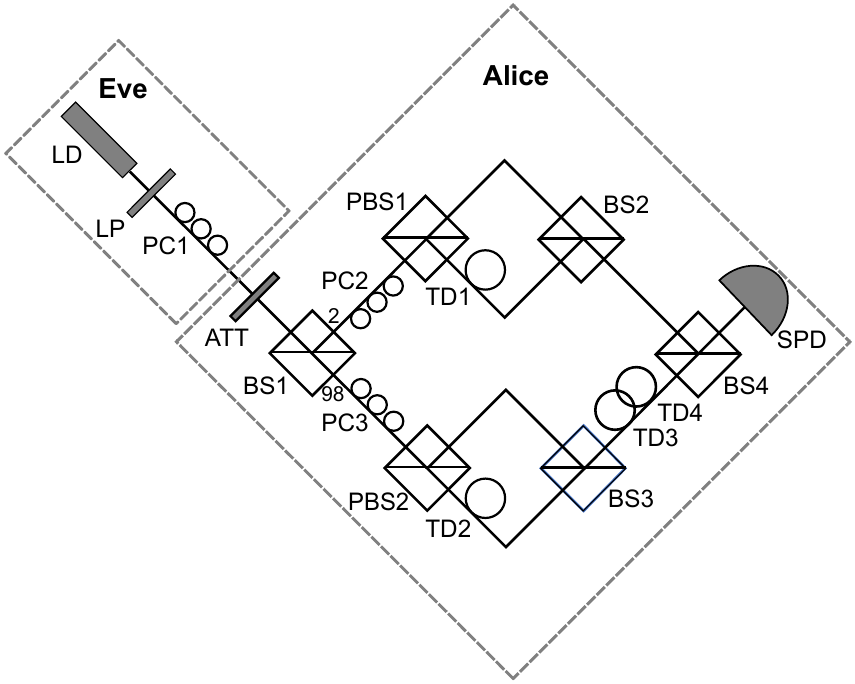}
    \caption{Experimental scheme for SI~QRNG (reprinted from Ref.~\onlinecite{cao2016}). The untrusted party Eve prepares quantum states and sends them to Alice, who is trusted. Alice then generates random numbers. Here, LD is a laser diode, LP is a linear polarizer, PC is a polarization controller, ATT is an optical attenuator, BS is a beam splitter, PBS is a polarization beam splitter, TD is a time delay, SPD is a single-photon detector.}
    \label{fig:SIQRNG}
\end{figure}

The experimental demonstration is shown in~\cref{fig:SIQRNG}. Weak coherent pulses are prepared with $\ket{+}$ polarization by a linear polarizer (LP) and a polarization controller (PC1). At Alice's side, a beam splitter (BS1) with splitting ratio $2$:$98$ is used to passively choose the $X$ or $Z$ basis. In~\cref{fig:SIQRNG}, the upper and lower paths correspond to the $X$ basis and $Z$ basis respectively. A single-photon detector is time-division-multiplexed by using four time delays TD1--TD4. For each coherent state Eve sends, a click in the first detection slot indicates that Alice chooses the $X$ basis and correctly detects the incoming pulse as $\ket{+}$, while a click in the second slot indicates a wrong detection, $\ket{-}$, which is used for the error estimation. Moreover, a click in the third slot indicates that Alice selects the $Z$ basis and obtain the result $\ket{H}$, while a click in the fourth slot indicates the result $\ket{V}$. 

\subsection{Hacking}

This SI~QRNG protocol assumes that the source can be untrusted, but the measurement station is trusted and characterized~\cite{cao2016}. However, it is not clear how to guarantee the latter requirement in practice. Therefore, Eve might be able to prepare fake states to generate nonrandom numbers. The detector blinding attack~(see~\cref{hack_ISS_QDS}) could force the SPD to work as a classical detector. Then Eve could send a strong bright pulse to trigger a detection in the first slot. Then she sends another bright pulse with either the state $\ket{H}$ or $\ket{V}$ to control the detection in the third or fourth slot. The attack can result in equal detection rates for $\ket{H}$ and $\ket{V}$, which looks like random clicks to Alice, while being precisely controlled by Eve. Eve actually thus controls the bit string.

Another potential issue is the wavelength-dependent attack, because the splitting ratio of a beam splitter might be sensitive to the wavelength of the incoming light~(see~\cref{hack_ISS_QDS}). All four beam splitters in the measurement station might be affected. By controlling the splitting ratio of BS1 and/or BS4, Eve can bias whether error checking or random bit generation happens. For BS2 and BS3, by manipulating the splitting ratio, Eve is able to partially control the results of error checking and bit generation. Please note that a wavelength filter alone will not protect the system from this attack, because Eve could send bright states to overcome the finite extinction ratio in the filter's stopband.

\section{Quantum secure direct communication}
\label{sec:QSDC}

QSDC transmits secret information directly through a quantum channel, instead of establishing a secret key first~\cite{deng2003}. The initial QSDC protocol is based on entangled pairs~\cite{wang2005c, wang2005d, wang2011a}. However, entanglement is not a necessary condition for QSDC. The first single-photon QSDC protocol, Deng-Long 2004 (DL04), was proposed in Ref.~\onlinecite{deng2004}. Recently, researchers started studying the strict security proof of this DL04 protocol~\cite{hu2017}. However, regarding the practical security, the implementation of this protocol also needs to be investigated. Also, more attention may need to be paid to the motivation for secure direct communication.

\subsection{Protocol and implementation}

The DL04 protocol contains two phases of channel estimation and a phase of secret transmission. The first channel estimation checks the security of the channel from Alice to Bob. Alice prepares a sequence of photons randomly chosen from the set of states $\ket{H}$, $\ket{V}$, $\ket{+}$, and $\ket{-}$, and sends them to Bob. He randomly selects a portion of the received photons, and randomly measures them in the $X$ or the $Z$ basis. Then Bob announces the measurement results and compares them to Alice's prepared states to calculate an error rate. Only when the error rate is lower than a threshold, Alice and Bob trust the channel and continue to the next step. Bob randomly selects another small portion of the received photons, and applies one of two unitary operations to each of them: $U = \ket{0}\bra{1}-\ket{1}\bra{0}$ or $I = \ket{0}\bra{0}+\ket{1}\bra{1}$, i.e., flipping a state or not. These photons are employed to check the security of the channel from Bob to Alice. The rest of the photons received by Bob are used to encode secret information by randomly applying the operator $U$ or $I$ to each photon. All these photons are then sent to Alice, who measures these photons in the preparation bases. Regarding the photons used for the security check, Alice checks if her measurement result is compatible with Bob's operation to estimate the error rate. Once the error rate is lower than a threshold, they trust the channel from Bob to Alice. The remaining photons measured in their preparation bases allow Alice to deterministically know Bob's operation, obtaining the secret information.

\begin{figure}
    \includegraphics[width=\columnwidth]{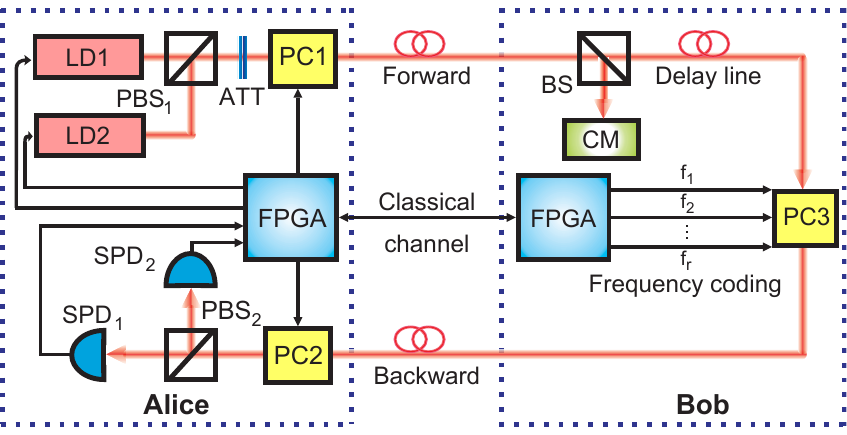}
    \caption{QSDC implementation (reprinted from Ref.~\onlinecite{hu2016}). Alice prepares and measures states, and Bob encodes the secret message by manipulating the states. Here, LD is a laser diode, PBS is a polarization beam splitter, ATT is an optical attenuator, PC is a polarization controller, BS is a beam splitter, CM is a control module, FPGA is a field programmable gate array, SPD is a single-photon detector.}
    \label{fig:QSDC}
\end{figure}

The protocol is implemented by the setup shown in~\cref{fig:QSDC}~\cite{hu2016}. Alice prepares the initial photon string and measures the photons encoded by Bob. She first prepares $\ket{H}$ and $\ket{V}$ using two laser diodes. The preparation and measurement bases are selected by $\rm{PC_1}$ and $\rm{PC_2}$ respectively. Bob encodes his information by $\rm{PC_3}$. All the basis choices are controlled by field programmable gate arrays~(FPGAs). The channel from Alice to Bob is denoted forward channel, and the channel from Bob to Alice is denoted backward channel. A beamsplitter at Bob's side selects a small portion of the received photons, and then a control module is used to check the security of the forward channel. The control module's scheme is the same as the passive measurement station in BB84 QKD system. A delay line is used to store the photons during the forward-channel check. To tolerate photon loss during secret transmission, a special method named single-photon frequency encoding is used. Instead of encoding information on individual photons, this method encodes information on the spectrum of a sequence of photons. After Alice detects a sequence of photons and converts them to a binary bit string, the spectrum can be known by applying the Fourier transform to the bit string. During the detection, Alice might miss some photons due to channel loss and imperfect detection efficiency. Fortunately, because the information does not only rely on an individual photon, but is determined by the spectrum of the entire sequence, missing some photons just reduces the signal-to-noise ratio, but the feature of the spectrum still exists~\cite{hu2016}. The calculated spectrum corresponds to the bit string that is the initial information Bob sent.

\subsection{Hacking}

The first phase of the DL04 QSDC protocol, the security check of the forward channel, is similar to the raw key exchange, sifting and error estimation phases in the BB84 QKD protocol~\cite{bennett1984}. The security check of the backward channel and secret direct transmission are quantum versions of the one-time pad, which randomly flips the bit information~\cite{deng2004}. Just as for QKD, the implementation~\cite{hu2016} may contain side channels.

The first potential side channel is that detectors may be attacked by the detector blinding attack~(see~\cref{hack_ISS_QDS}). During the check of the forward channel, Eve blinds the detectors in the control module and conducts an attack with fake states~\cite{makarov2005} to control Bob's detection results. Since this attack introduces no errors, the security check is passed. During the second check of the backward channel and information transmission, Eve uses classical optical detectors to measure her bright pulses modulated by Bob. Since these are states resent by Eve during the previous phase, Eve could apply the same basis as in the previous step to know with certainty what operation Bob performed. Then, she sends the same states with proper brightness to Alice's blinded detectors, such that only when Alice selects the same bases as Eve, Alice obtains detections. This attack results in full control of Alice’s measurement outcomes. Again, no extra errors are introduced. Furthermore, Eve learns the secret information between Alice and Bob. This breaks the security of QSDC. Please note that because this implementation uses an active basis choice (the basis is actively selected by the polarization controller), Eve's measurement basis can only match Alice's/Bob's measurement basis half the time. However, when the basis matches the click probability in Bob under attack can be unity, while his single-photon detection efficiency is typically much lower than unity~\cite{lydersen2010a}. This may compensate for the extra loss introduced by the attack.

The second possible side channel exists in the polarization controllers that might be vulnerable to the Trojan-horse attack~(see~\cref{hack_DSS_QDS}). In this QSDC implementation~\cite{hu2016}, Eve can conduct the Trojan-horse attack on $\rm{PC_1}$ or $\rm{PC_3}$. From an attack on $\rm{PC_1}$, Eve would know Alice's basis choice in the state preparation and measurement, as $\rm{PC_2}$ applies the same basis as $\rm{PC_1}$. The difference between the prepared and the measured state is Bob's secret information~(flip or not). On the other hand, Bob's encoded information could be directly known by hacking $\rm{PC_3}$~(similarly to~\cref{two_way_trojan}). Once Eve knows the original states prepared by Alice or what Bob’s modulation was, she could obtain the secret information.

\section{Blind quantum computing}
\label{sec:BQC}

In the future, a quantum computer could be used as a server that provides quantum computation capability to remote users, who themselves do not have a quantum computer and only use simple technology. A key task is to keep the client's data and program secret from the server. Classical blind computing protocols exist~\cite{rivest1978a}, but it can only guarantee computational security~\cite{barz2012}. However, taking advantage of quantum mechanics, BQC is able to provide unconditional security for client's data and computation in the quantum computer server~\cite{broadbent2009}.

\subsection{Protocol and implementation}

BQC is based on entangled multiparticle cluster states~\cite{barz2012}. In the BQC protocol, qubits are first prepared as $\ket{\theta_j} = (\ket{0} + e^{i\theta_j}\ket{1})/\sqrt{2}$ by a client, where $\theta_j$ is randomly selected from \{$0, \pi/4, ..., 7\pi/4$\}. Then the single-photon qubits are sent to a quantum server that entangles them with each other by applying controlled-phase gates, so that the qubits form a cluster state. Then the cluster state is measured by the quantum server, which performs single-qubit measurements in the basis $\ket{\pm_{\delta_j}} = (\ket{0} + e^{i\delta_j}\ket{1})/\sqrt{2}$. The measurement basis is instructed by the client: $\delta_j = \phi_j + \theta_j + \pi r_j$, where $\phi_j$ is the desired rotation and $r_j$ is randomly chosen from \{0, 1\}. Since $\theta_j$ is the initial phase hidden from the quantum server, the server is not able to calculate the desired rotation $\phi_j$  from the measurement result. It is remarkable that for the cluster state, its shape, such as the dimension, also may leak information about the operation gates. Thus, also the shape of the cluster state is required to be hidden, which can be accomplished by choosing, for example, brickwork states~\cite{barz2012}. The BQC protocol then completes a quantum computation while preserving the client's privacy.

\begin{figure}
    \includegraphics{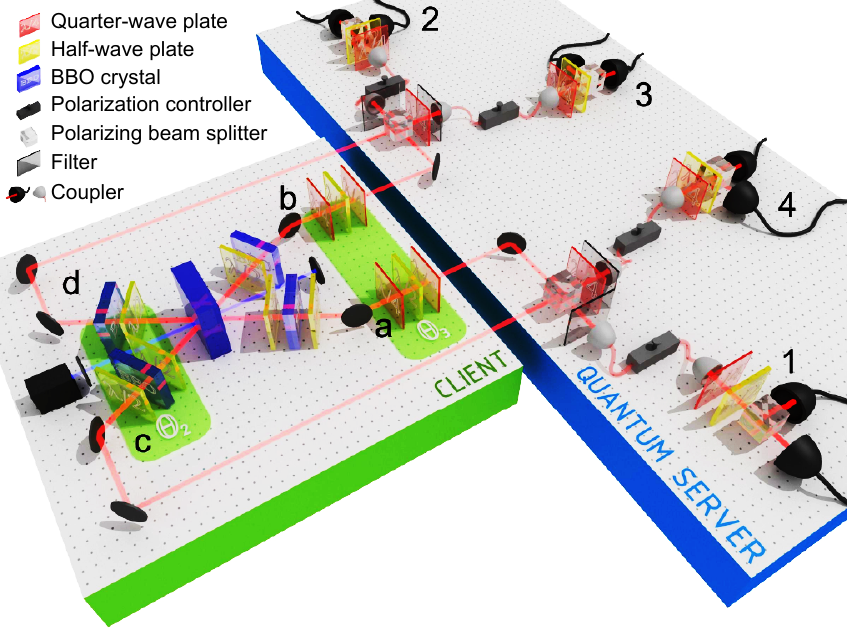}
    \caption{Proof-of-principle implementation of BQC (reprinted from Ref.~\onlinecite{barz2012}). Entangled photon pairs are generated from non-collinear type-II spontaneous
parametric down-conversion process in BBO crystal. The horizontal and vertical polarization represents $\ket{0}$ and~$\ket{1}$.}
    \label{fig:BQC}
\end{figure}

Theoretically, the client only needs to have a single-photon source to generate a state $\ket{\theta_j}$ and send it to the server. However, implementing a single-photon source is challenging so far, as standard parametric down-conversion sources always also have higher-order emissions, meaning that instead of one pair, two or more pairs are emitted at the same time. An initial demonstration of the BQC protocol with current technology is shown in~\cref{fig:BQC}~\cite{barz2012}. Note that in the current implementation, entangled pairs are first prepared on the client's side, and the cluster state is generated on the server's side. The laser beam passes a BBO crystal to first generate the entangled pair traveling forwards. Then the beam is reflected and passes the BBO crystal again to generate the entangled pair traveling backwards. The initial phase $\theta_j$ is applied by rotating the angles of half-wave plates and quarter-wave plates, serving as modulators. Then the entangled states are sent to the quantum server's side, where a cluster state is generated. The states are measured in different bases, as instructed by the client. In this BQC protocol, the setup of the client is relatively simple, but the setup of the quantum server would have more capabilities once a real quantum computer is available. Here we take one type of BQC protocol as an example. There also exist other versions of BQC where the server generates entangled cluster states and the measurements are done by the client~\cite{morimae2013,greganti2016}.

\subsection{Hacking}

In the above subsection, a proof-of-principle implementation of BQC was introduced. Although in the future the technology available to implement the quantum server for BQC will be more mature and comprehensive, the client setup is already relatively clear. It is foreseeable that future client station will likely still consist of a photon source and modulators. Unfortunately, in practice, any kind of modulator is susceptible to the Trojan-horse attack~\cite{vakhitov2001,gisin2006, jain2014, sajeed2015}. This vulnerability breaks an important assumption in BQC: the initial phase $\theta_j$ should be unknown to the untrusted quantum server. Specifically, regarding the implementation shown in~\cref{fig:BQC}, the phase modulation is done by the wave plates. The reflected light from the wave plates may leak information about $\theta_j$. Instead of wave plates, an advanced setup in the future could be using phase modulators to randomly modulate the phase $\theta_j$, which is a technique widely used in quantum cryptography~\cite{takesue2005,bogdanski2008,collins2016}. Unfortunately, the Trojan-horse attack might still be applied to phase modulators, as we have discussed in~\cref{hack_DSS_QDS}. 

Except for imperfect phase modulation, another possible issue is the photon source itself. For the current version of implementation, the entanglement source sometimes might simultaneously emit multiple pairs of entangled states. In this case, Eve could split off a copy of entangled states from the source. Then measuring her copy would give Eve information about the state itself. Even in future implementations, when ideal single-photon sources are available, one still needs to pay attention to state generation. For instance, the BQC protocol needs indistinguishable multiple photons~\cite{barz2012}. Thus, careful source design is crucial to avoid any distinguishability in the generated photons (this can, in principle, occur in any degree of freedom, for example wavelength).

For other variations of BQC protocols, where the measurements are done on the client's side~\cite{morimae2013,greganti2016}, attacks that leak information about the measurement settings are applicable. So, in a setting where the client uses wave plates to choose a measurement basis~\cite{greganti2016}, the Trojan-horse attack could be applied as well.

\section{Countermeasures}
\label{Countermeasure}

An imperfect implementation compromises the security promised in theory, as we have argued in~\cref{sec:QDS,sec:QSS,sec:QRNG,sec:QSDC,sec:BQC}. To patch the practical loopholes, we should consider feasible countermeasures in implementations of quantum cryptographic protocols. Existing countermeasures for QKD and countermeasures under development may be adaptable to implementations of other cryptographic protocols. However, integrating these considerations into the relevant security proofs is an open challenge. We now recap countermeasures proposed in the literature for both the source and measurement parts of a quantum cryptographic system. We also discuss how they may be applied to the protocols surveyed in this Article.

\subsection{Countermeasures against source imperfection}

\begin{figure}
    \includegraphics{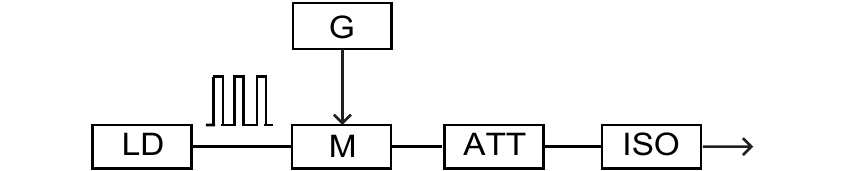}
    \caption{The scheme of countermeasures against source imperfections. To eliminate the mismatch among different laser diodes, a single laser diode LD followed by a modulator~M can be used. The following attenuator ATT and isolator ISO provide sufficient isolation to prevent the Trojan-horse attack in a unidirectional system. Here, G denotes a modulation signal generator.}
    \label{fig:source}
\end{figure}

Properly implementing the quantum-state source in the above protocols requires that any other degrees of freedom are uncorrelated with the degree of freedom where information is encoded. However, for the states prepared by different laser diodes~(see~\cref{ISS_QDS}), the laser diodes may show the inherent difference in the spectrum and emission time. These types of difference hint which laser diode is on, i.e.,\ which state is prepared. The mismatch in a certain degree of freedom could be a side channel for Eve who tries to distinguish different quantum states~\cite{nauerth2009,huang2018}. To avoid this inherent mismatch among different laser diodes, quantum state preparation could use only one laser diode followed by optical modulators~(\cref{fig:source}), as shown in many QKD implementations~\cite{stucki2002,stucki2009,tang2014}. The laser diode generates identical pulses. Then different states are modulated by a phase modulator~\cite{stucki2002}, intensity modulator~\cite{stucki2009}, or polarization modulator~\cite{tang2014}.

The external modulation method could be applied to the implementation of double-receiver QDS~(see~\cref{ISS_QDS}). However, this modification might open another loophole: the Trojan-horse attack on the modulators. Once a system uses a modulator, countermeasures against Trojan-horse attack are required. For a unidirectional system that only sends states from one party to another but never back, a possible countermeasure is adding enough isolation between the modulator and the output port connected to the quantum channel, as shown in~\cref{fig:source}. The amount of isolation is defined by the combination of bidirectional attenuation from attenuators, the unidirectional attenuation from isolators and total reflection probability from lasers and modulators. For example, in a BB84 QKD system, the isolation has been quantified as the following~\cite{lucamarini2015}. Suppose Eve injects pulsed light into the party preparing the state. The injected power is limited by the maximum power transmitted safely through standard single-mode fiber (assumed to be 12.8~W in Ref.~\onlinecite{lucamarini2015}). The amount of reflection then is obtained after the injected light is attenuated by the system isolation. Taking this amount of reflection into account in the calculation of the key rate, one could obtain the final secure key rate. To obtain a key rate under the Trojan-horse attack that is close to the rate without an attack, 170 dB isolation is required~\cite{lucamarini2015}.

A similar methodology could be applied to single-receiver QDS, QSDC, and BQC, which may be vulnerable to the Trojan-horse attack. In each implementation, attenuators and isolators could be added between the modulators and system output, and the reflectivity of modulator and laser diodes should be quantified. Then the required amount of isolation should be calculated according to the security models of the corresponding protocol as has already been done for QKD~\cite{lucamarini2015}. The chosen amount of isolation should maintain the system's security properties. We notice that in the implementation of single-receiver QDS in Ref~\onlinecite{amiri2016}, an attenuator is already included to weaken the output power to single-photon level. However, this amount of attenuation is probably not sufficient to provide isolation to counter the Trojan-horse attack.

For a bidirectional plug \& play QSS system (see~\cref{QSS_single}) and pass-through QSDC (see~\cref{sec:QSDC}), the system's isolation in the previous countermeasure is not applicable, because it would block transmission of the states. In the bidirectional system, single-photon monitors would be needed to observe the incoming light~\cite{vakhitov2001}. It is not clear if implementing such countermeasure securely is realistic. Nevertheless, a patent by Trifonov and Vig~\cite{trifonov2007} proposes a scheme against Trojan-horse attack with a single-photon watchdog detector. This countermeasure could be adapted for the single-qubit QSS implementation. Alice could employ a watchdog detector for the received light. The rest of parties in the scheme could use two watchdog detectors to observe two fiber connection ports at each side of the PM. Any alarm would abort the protocol. Please note that the single-photon detector might be vulnerable to the detector blinding attack. Thus, a corresponding countermeasure against detector control attacks is necessary, which is discussed in the next subsection.

\subsection{Countermeasures against measurement imperfection}

In a party that makes measurements, characteristics of passive optical components, such as beam splitters, might be sensitive to wavelength. That is, the component's behavior for unexpected wavelengths may deviate from what is assumed. To provide practical security, wavelength dependence should be eliminated. A possible method is using a wavelength filter to block unexpected wavelengths, and only pass a narrow range around the working wavelength~\cite{lucamarini2015}. In the implementation of double-receiver QDS (see~\cref{ISS_QDS}), this filter could be added before the beam splitter in Bob and Charlie, i.e.,\ right at their input ports. The filter's transmission should be verified in a wide range of wavelengths. However, there is a limitation to this approach: Eve can simply increase her light power to pass through the stopband. Therefore, as a more robust countermeasure, we suggest utilizing active basis choice in the measurement station.

Another major vulnerability in measurement setups is imperfections in single-photon detectors (see~\cref{hack_ISS_QDS}). A proposed countermeasure for QKD systems is calibrating the characteristics of detectors in real time, avoiding Eve's manipulation~\cite{maroy2016}. In this receiver design, a calibrated light source is locally included in the measurement unit, in combination with several other countermeasures. By randomly activating this local source to send photons to the detectors, the corresponding detection efficiency can be calibrated during the system operation. The characterized detection efficiency can then be used in the security proof to calculate the secure key rate. A similar design might be applicable to measurement stations in the other quantum cryptographic protocols. However, incorporating the calibration procedure into their security models should be studied in each case.

Another approach to entirely avoid the effect of imperfect detectors and other measurement imperfections are measurement-device-independent (MDI) quantum cryptographic protocols~\cite{fu2015}, such as MDI QKD~\cite{lo2012}, MDI QSS~\cite{yang2016} and MDI QDS~\cite{puthoor2016, yin2017}. In the MDI protocols, the party making measurements is untrusted: there are no security assumptions regarding the measurements. Even if Eve makes the measurements, the secret information (provided the protocol produces it) can still be distributed among the rest of the authenticated parties. This is a promising idea to avoid security loopholes related to the measurements. However, state preparation remains trusted and still needs to be carefully designed to avoid loopholes.

\section{Conclusion}
\label{conclusion}

We have surveyed implementations of five types of quantum cryptographic primitives. As our analysis shows, these quantum cryptographic systems might have security loopholes similar to QKD systems, because they use similar optical components. These imperfections would compromise the security properties of each quantum cryptographic protocol (see summary in~\cref{tbl:summary}). We discuss implementations of these protocols, showing that practical insecurity is a common issue in the implementation of quantum cryptography in general, not only in QKD. In other words, a gap between perfect theory and imperfect practice generally exists in quantum cryptography. 

Our analysis of imperfections in this survey has been intended to reveal a broad picture. Detailed analysis of imperfections for each specific implementation should be done in the future. Once the existence of practical loopholes has been noticed, it becomes essential to bridge the gap between theory and practice. One should consider countermeasures when implementing existing protocols or designing new quantum cryptographic protocols that tolerate practical imperfections. Fortunately, these approaches appear to be feasible. However, integrating imperfections into security proofs~\cite{ma2005,lucamarini2015,tamaki2016} is a significant challenge, which should be addressed in future studies.

\begin{acknowledgments}
We thank G.~Brassard for discussions. We acknowledge financial support from NSERC of Canada (programs Discovery and CryptoWorks21) and Ontario MRIS. A.H.\ acknowledges support by China Scholarship Council. S.B.\ acknowledges support from the Carl Zeiss Foundation. E.A.\ acknowledges support from the \nohyphens{EPSRC} UK Quantum Communication Hub (grant number \nohyphens{EP/M013472/1}).
\end{acknowledgments}

\def\bibsection{\medskip\begin{center}\rule{0.5\columnwidth}{.8pt}\end{center}\medskip} 

\end{document}